\documentstyle[12pt,a4,epsf]{article}

\title{{\bf An improved gamma-ray limit on the density of
 primordial black holes}}
\author{A. Barrau, G. Boudoul, L. Derome}
\date{{\small Laboratoire de Physique Subatomique et de Cosmologie, CNRS-IN2P3/UJF\\
 53, av des Martyrs, 38026 Grenoble Cedex, FRANCE\\
 }}

\begin{document}

\bigskip

\maketitle

\bigskip

\begin{center}
{\bf Abstract}\\
\end{center}
Gamma-rays are, with antiprotons, a very efficient way to derive upper limits on
the density of evaporating black holes. They have been successfully used in the
last decades to severely constrain the amount of Primordial Black Holes (PBHs) in our Universe. This
article suggests a little refinement, based on the expected background, to improve
this limit by a factor of three. The resulting value is :
$\Omega_{PBH}<3.3\times 10^{-9}$. 
\\

%\begin{flushleft}
%{\it PACS}: 97.60.Lf, 98.70.Sa\\
%{\it Keywords}: Primordial black holes
%\end{flushleft}

\bigskip

\newpage

Small black holes could have formed in the early Universe if the density
contrast was high enough (typically $\delta > 0.3 - 0.7$ , depending on models). 
Since it was discovered by Hawking \cite{Hawking5} that they should evaporate with
a black-body like spectrum of temperature $T=\hbar c^3/(8\pi k G M)$, the emitted
cosmic rays have been considered as the natural way, if any, to
detect them. Those with initial masses smaller than $M_*\approx 5\times
10^{14}~{\rm g}$ should have finished their evaporation by now whereas those with
masses greater than a few times $M_*$ do emit nothing but extremely low energy
massless fields. The emission spectrum for
particles of energy $Q$ per unit of time $t$ is, for each degree of
freedom, given by :
\begin{equation}
\frac{{\rm d}^2N}{{\rm d}Q{\rm
d}t}=\frac{\Gamma_s}{h\left(\exp\left(\frac{Q}{h\kappa/4\pi^2c}\right)-(-1)^{2s}\right)}
\end{equation}
where $\kappa$ is the surface  gravity, $s$ is the
spin of the emitted species and $\Gamma_s$ is the absorption probability 
proportional to $M^2Q^2$ in the high energy limit (contributions of angular 
velocity and electric potential have been neglected since the
black hole discharges and finishes its rotation much faster than it
evaporates). PBHs have been investigated for many different purposes, including
tests for quantum gravity \cite{khlo} that are especially active nowadays.\\

As was shown by MacGibbon and Webber \cite{MacGibbon1}, when the
black hole temperature is greater than the quantum chromodynamics confinement 
scale $\Lambda_{QCD}$, quark and gluon jets are emitted instead of composite 
hadrons. This should be taken into account when computing the cosmic-ray flux
expected from their evaporation. Among all the emitted particles, two species are
especially interesting : gamma-rays around 100 MeV because the Universe is very
transparent to those wavelengths and because the flux from PBHs becomes softer 
($\propto E^{-3}$ instead of $\propto E^{-1}$)
above this energy, and antiprotons around 0.1-1 GeV \cite{aurel} because the natural background due to
spallation of protons and helium nuclei on the interstellar medium is very
small and fairly well known. This article aims at taking into account the
contribution from blazars as well as from normal galaxies in the gamma-ray
background to reduce the available window for PBHs.\\

Computing the contributions both from the direct electromagnetic emission
and from the major component resulting from the decay of neutral pions, the
gamma-ray spectrum from a given distribution of PBHs can be compared with 
measurements. 

The flux on Earth can be written as:

$$\frac{d^2\Phi}{dEdt}=\frac{1}{2}\int_{t_{form}}^{t_0}\left(
\frac{R(t)}{R_{form}}\right)e^{-\tau(t,E)}\int_{M_*(t)}^{\infty} 
\frac{d^2 \phi}{dEdt}(M(t,M_i),E'=E\frac{R_0}{R})\frac{d^2n}{dM_idV}dM_icdt
$$
where $t_{form}$ is the formation time, $t_0$ is the age of the Universe,
$\tau$ is the optical depth, $R(t)$ is the scale factor of the Universe at time 
$t$, $\phi$ is the individual gamma spectrum from a PBH and $d^2n/dM_idV$ is
the initial mass spectrum. When compared with the observations, 
this translates (in units of critical density) 
into \cite{Carr1998}~: $\Omega_{PBH}(M_*) < 1.0\times 10^{-8}$, bettering substantially previous 
estimates \cite{MacGibbon1991}.\\

This limit can be improved when taking into account the "guaranteed" gamma-ray
background. Computing the contribution from unresolved blazars and the emission
from normal galaxies, Pavlidou \& Fields \cite{pav} have estimated the
minimum amount of extragalactic gamma-rays which should be expected. The first
one was computed using the Stecker-Salamon model and the second one is assumed
to be proportional to the massive star formation rate (which is itself
proportional to the supernovae explosion rate) as it is
due to cosmic-ray interactions with diffuse gas. This background at 100 MeV is
estimated at $\Phi_{TH}=5.45\times 10^{-14}~{\rm cm}^{-3}{\rm GeV}^{-1}$. Using
the Carr \& MacGibbon estimation at the same energy \cite{Carr1998} 
$\Phi_{PBH}=7.5\times 10^{-6}\Omega_{PBH}~{\rm cm}^{-3}{\rm GeV}^{-1}$, a new limit can be
obtained by requiring that $\Phi_{PBH}+\Phi_{TH}<\Phi_{EGRET}$ where
$\Phi_{EGRET}$ is the measured flux \cite{EGRET}. To evaluate this later in a
very conservative way, both the normalisation and the spectral index were chosen
(within the error bars) in this paper at the value leading to the highest 100
MeV flux : $\Phi_{EGRET} < 7.94\times 10^{-14}~{\rm cm}^{-3}{\rm GeV}^{-1}$.
This leads to $\Omega_{PBH}(M_*)<3.3\times 10^{-9}$.\\

From the cosmological point of view, this new limit improves directly the
estimates of the maximum allowed PBH mass fraction $\beta$:
$$
\label{beta}
  \beta(M_{H})=\frac{1}{\sqrt{2\pi}\,\sigma_H(t_{k})}~
           \int_{\delta_{min}}^{\delta_{max}}\,
           e^{-\frac{\delta^2}{2 \sigma_H^2(t_{k})}}\,\textrm{d}\delta
          \approx\frac{\sigma_H(t_{k})}{\sqrt{2\pi}\,\delta_{min}}
           e^{-\frac{\delta_{min}^2}{2 \sigma_H^2(t_{k})}}~,
$$
where $t_k$ is the horizon crossing time for the considered mode, $\delta$ the
density contrast ($\delta_{min}\approx 0.7$), $M_H$ is the Hubble mass at $t_k$ and $\sigma_H^2(t_k)\equiv
\sigma^2(R)|_{t_k}$ where $\sigma^2(R) 
\equiv <(\frac{\delta M}{M})^2_R>$ 
is computed with a filtering window function with
$R=\frac{H^{-1}}{a}|_{t_k}$. The latest computations \cite{po} relate this
value to the density parameter by
$$\Omega_{PBH}(M)h^2=6.35\times 10^{16}\times \beta(M)\left( \frac{10^{15}g}{M}
\right).$$ Our new limit leads to $\beta(M_*)<1.3\times 10^{-26}$ which is
compatible with antiprotons estimations \cite{barrau} and remains the only
observational access to such small scales in the early Universe.\\

Measurements from the GLAST satellite and more refined theoretical predictions
on the background could slightly improve these results but both gamma-rays and
antiprotons seem to have closed their detection windows. One of the last hopes
could reside in antideuterons which are very rarely induced by spallation
below 1 GeV for kinematical reasons \cite{barrau2}.

\end{document}